\documentclass[11pt,twoside]{article}

\usepackage{asp2014}

\aspSuppressVolSlug
\resetcounters

\begin{document}

\title{CAESAR: Space Weather archive prototype for ASPIS}

\author{
Marco~Molinaro,$^1$ Valerio~Formato,$^2$ Carmelo~Magnafico,$^3$ Federico Benvenuto,$^4$ Alessandro~Perfetti,$^3$ Rossana~De~Marco,$^3$ Cristina~Campi,$^4$ Andrea~Tacchino,$^4$ Valeria~di~Felice,$^2$ Ermanno~Pietropaolo,$^5$ Giancarlo de~Gasperis,$^6$ Luca~di~Fino,$^6$ Gregoire~Francisco,$^6$ Igor~Bertello,$^3$ Anna Milillo,$^3$ Giuseppe~Sindoni,$^7$ Christina~Plainaki,$^7$ Marco~Giardino,$^8$ Gianluca~Polenta,$^8$ Dario~Del~Moro,$^6$ and Monica~Laurenza$^3$
}
\affil{$^1$INAF - Astronomical Observatory of Trieste, Trieste, Italy; \email{marco.molinaro@inaf.it}}
\affil{$^2$INFN - Rome Tor Vergata Section, Rome, Italy}
\affil{$^3$INAF - Institute of Space Astrophysics and Planetology, Rome, Italy}
\affil{$^4$University of Genova - Department of Mathematics, Genova, Italy}
\affil{$^5$University of L'Aquila - Department of Physical and Chemical Sciences, L'Aquila, Italy}
\affil{$^6$University of Rome ``Tor Vergata'' - Department of Physics, Rome, Italy}
\affil{$^7$Agenzia Spaziale Italiana (ASI), Rome, Italy}
\affil{$^8$ASI - Space Science Data Center, Rome, Italy}

\paperauthor{Marco~Molinaro}{marco.molinaro@inaf.it}{0000-0001-5028-6041}{INAF}{Osservatorio Astronomico di Trieste}{Trieste}{}{}{Italy}
\paperauthor{Valerio~Formato}{valerio.formato@roma2.infn.it}{0000-0002-8921-3832}{INFN}{Sezione di Roma Tor Vergata}{Rome}{}{}{Italy}
\paperauthor{Carmelo~Magnafico}{carmelo.magnafico@inaf.it}{0000-0001-5066-0267}{INAF}{Istituto di Astrofisica e Planetologia Spaziali}{Roma}{}{}{Italy}
\paperauthor{Federico~Benvenuto}{benvenuto@dima.unige.it}{}{Universit\'{a} di Genova}{Dipertimento di Matematica}{Genova}{}{}{Italy}
\paperauthor{Alessandro~Perfetti}{alessandro.perfetti@inaf.it}{0000-0002-4014-5701}{INAF}{Istituto di Astrofisica e Planetologia Spaziali}{Roma}{}{}{Italy}
\paperauthor{Rossana~De~Marco}{rossana.demarco@inaf.it}{0000-0002-7426-7379}{INAF}{Istituto di Astrofisica e Planetologia Spaziali}{Roma}{}{}{Italy}
\paperauthor{Cristina~Campi}{campi@dima.unige.it}{}{Universit\'{a} di Genova}{Dipertimento di Matematica}{Genova}{}{}{Italy}
\paperauthor{Andrea~Tacchino}{a.tacchino@gmail.com}{}{Universit\'{a} di Genova}{Dipertimento di Matematica}{Genova}{}{}{Italy}
\paperauthor{Valeria~di~Felice}{valeria.difelice@roma2.infn.it}{}{INFN}{Sezione di Roma Tor Vergata}{Rome}{}{}{Italy}
\paperauthor{Ermanno~Pietropaolo}{ermanno.pietropaolo@aquila.infn.it}{0000-0002-6633-9846}{Universit\'{a} dell'Aquila}{Dipartimento di Scienze Fisiche e Chimiche}{L'Aquila}{}{}{Italy}
\paperauthor{Giancarlo~de~Gasperis}{giancarlo.degasperis@roma2.infn.it}{}{Universit\'{a} di Roma ``Tor Vergata''}{Dipartimento di Fisica}{Rome}{}{}{Italy}
\paperauthor{Luca~di~Fino}{luca.difino@roma2.infn.it}{}{Universit\'{a} di Roma ``Tor Vergata''}{Dipartimento di Fisica}{Rome}{}{}{Italy}
\paperauthor{Gregoire~Francisco}{gregoirefrancisco@gmail.com}{}{Universit\'{a} di Roma ``Tor Vergata''}{Dipartimento di Fisica}{Rome}{}{}{Italy}
\paperauthor{Igor~Bertello}{igor.bertello@inaf.it}{}{INAF}{Istituto di Astrofisica e Planetologia Spaziali}{Roma}{}{}{Italy}
\paperauthor{Anna~Milillo}{anna.milillo@inaf.it}{0000-0002-0266-2556}{INAF}{Istituto di Astrofisica e Planetologia Spaziali}{Roma}{}{}{Italy}
\paperauthor{Giuseppe~Sindoni}{giuseppe.sindoni@asi.it}{0000-0002-3348-7930}{ASI}{}{Rome}{}{}{Italy}
\paperauthor{Christina~Plainaki}{christina.plainaki@asi.it}{}{ASI}{}{Rome}{}{}{Italy}
\paperauthor{Marco~Giardino}{marco.giardino@ssdc.asi.it}{0000-0002-6983-6346}{ASI}{Space Science Data Center}{Rome}{}{}{Italy}
\paperauthor{Gianluca~Polenta}{gianluca.polenta@ssdc.asi.it}{0000-0002-6983-6346}{ASI}{Space Science Data Center}{Rome}{}{}{Italy}
\paperauthor{Dario~Del~Moro}{dario.delmoro@roma2.infn.it}{0000-0003-2500-5054}{Universit\'{a} di Roma ``Tor Vergata''}{Dipartimento di Fisica}{Rome}{}{}{Italy}
\paperauthor{Monica~Laurenza}{monica.laurenza@inaf.it}{0000-0001-5481-4534}{INAF}{Istituto di Astrofisica e Planetologia Spaziali}{Roma}{}{}{Italy}



\begin{abstract}
The project CAESAR (Comprehensive spAce wEather Studies for the ASPIS prototype Realization) is aimed to tackle all the relevant aspects of Space Weather (SWE) and realize the prototype of the scientific data centre for Space Weather of the Italian Space Agency (ASI) called ASPIS (ASI SPace Weather InfraStructure). This contribution is meant to bring attention upon the first steps in the development of the CAESAR prototype for ASPIS 
and will focus on the activities of the Node 2000 of CAESAR, the set of Work Packages dedicated to the technical design and implementation of the CAESAR ASPIS archive prototype.
The product specifications of the intended resources that will form the archive, functional and system requirements gathered as first steps to seed the design of the prototype infrastructure, and evaluation of existing frameworks, tools and standards, will be presented as well as the status of the project in its initial stage.
\end{abstract}



\section{Introduction}
\label{intro}

The Italian Space Agency (ASI) has produced \citep{2020JSWSC..10....6P} a Space Weather (SWE) roadmap 
for a long-term strategy to support the future scientific research of SWE, and the development of a related
national scientific data centre for Space Weather, called ASI SPace Weather
InfraStructure (ASPIS), while reinforcing the interactions and synergies among the SWE and Planetary 
Space Weather (PSW) Italian groups, and organizing them in a strong collaborative environment. 
The CAESAR\footnote{\url{https://caesar.iaps.inaf.it/}} \citep{2022Laurenza} project was selected
to reach such goals. CAESAR is coordinated by the National Institute of Astrophysics (INAF) and the project's 
partners span two more national institutes (INGV \& INFN) and seven universities (Tor Vergata -- Rome II, 
Perugia, Genova, L’Aquila, Calabria, Catania, Trento).
CAESAR will focus on the whole chain of phenomena from the Sun to the Earth up to planetary environments, 
selecting a number of well observed \textit{target Space Weather events} for detailed and comprehensive studies in
order to showcase the proposed approach.

\articlefigure[width=1.0\textwidth]{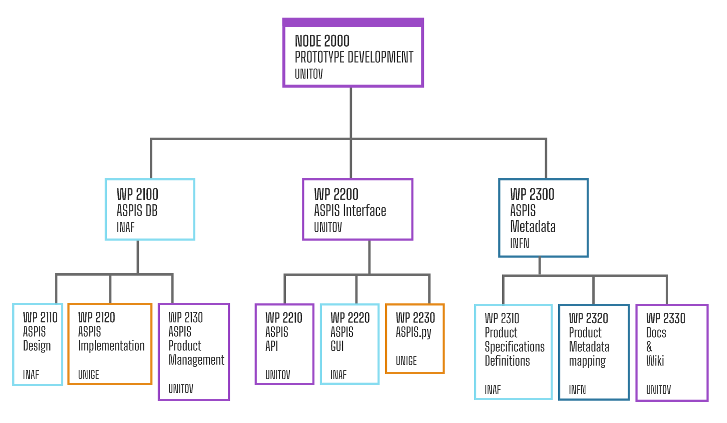}{p22_f1_node2000}{CAESAR Node 2000 WBS.}

The archive prototype itself will be designed and developed by the \textit{Node 2000} (see 
Fig.~\ref{p22_f1_node2000}) group of the CAESAR's Work
Packages (WPs), i.e. three high level WPs (they will be described by their \textit{pillar} activities in 
Sec.~\ref{prototype}) that will provide the database, the interfaces on top of it and the metadata it all 
relies upon.

\section{ASPIS prototype}
\label{prototype}

The CAESAR ASPIS prototype implementation will be hinged on three main pillars (their interconnection and connection to scientific project's WPs can be seen in Fig.~\ref{p22_f2_hldiag}):
\begin{description}
    \item[Database] (DB) managed by WP2100, taking care of the design and implementation of the DB as well as 
    the ingestion process;
    \item[Interface] managed by WP2200, that will define the API sitting on top of the DB as well as a web GUI 
    for basic usage of the archive and a, more advanced, python module (\textit{ASPIS.py}) to consume the 
    archive's content;
    \item[Metadata] managed by WP2300, that, besides taking care of the documentation, is devoted to collect and 
    map all data, metadata and products contributed to CAESAR by the data providers among its partners.
\end{description}

\articlefigure[width=1.0\textwidth]{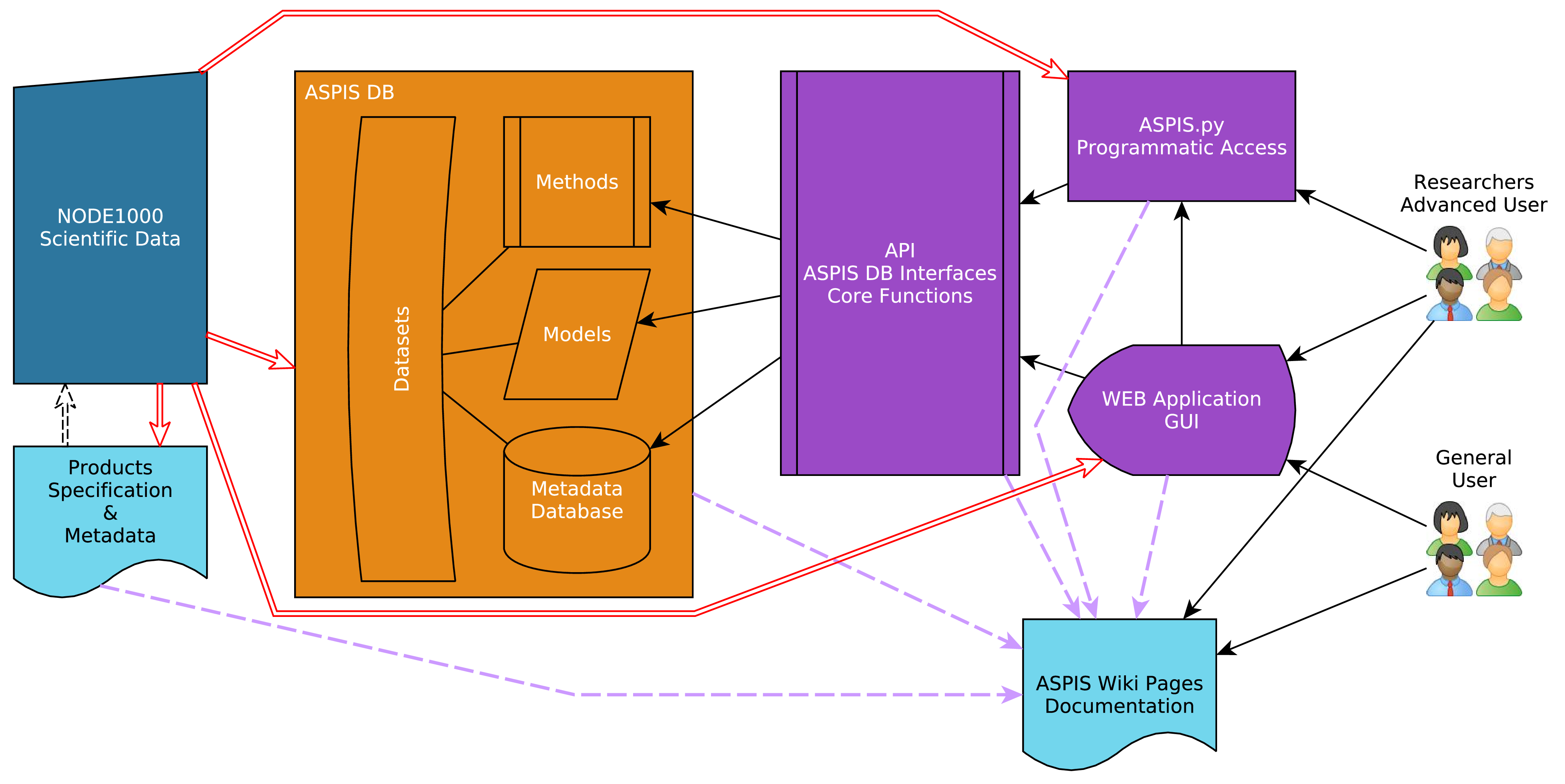}{p22_f2_hldiag}{High Level architecture of the CAESAR ASPIS 
prototype. Solid lines show access paths; double ones indicate requirements provided by the research community 
(and double dashed the provided template for product description); dashed lines show collection of documentation.}

The ASPIS DB will contain mainly proprietary/co-proprietary products, with their relative data policy in a
homogeneous, standardised collection of resources. Other important external data shall be accessed through links 
to existent archives. A set of 75 main products were identified in the proposal, spanning calibrated data, 
derived ones, models and tools.
Currently collected sample data and descriptions cover about a half of the above, for a, finer grained, current
total of more than 100 among data collections, data models and tools. The DB status is in its \textit{pilot} 
phase, with a relational database structure that has been defined and is about to be populated using sample data 
and metadata from the collected products.
To collect products metadata specifications and sample datasets from the contributed products there was a 
requirement to have both:
\begin{itemize}
    \item a way to save product metadata information in a machine useful format;
    \item a solution that suited scientific researchers without getting lost in technicalities.
\end{itemize}
For this reason a tool based on JSONForms\footnote{\url{https://jsonforms.io/}} 
(see next Sec.~\ref{prospect}) 
has been developed.

\section{Product Specifications}
\label{prospect}

\textit{ProSpecT} (\textit{Pro}duct \textit{Spec}ification 
\textit{T}emplate\footnote{\url{http://prospect-caesar.ssdc.asi.it/}})
is a tool that leverages on JSON Schema technology and JSONForms to provide a web interface to 
describe a ``Product'' (i.e. a \textit{Resource} in general web jargon) and generate a JSON object 
that contains all the required metadata.
The metadata schema (i.e. \textit{template} here) is written in JSON and is used, referenced by 
JSONForms elements, to generate the web form to be filled in by product providers. The output of 
the form is the desired JSON metadata instance describing the relevant product.
The metadata content has been mimicked from the IVOA 
VOResource\citep[][and its extensions]{2018ivoa.spec.0625P} standard adapted to the needs and
peculiarities of the CAESAR project.
The collected JSON descriptions of the about 100 products have already been used to generate 
wiki-like pages for documentation and to inspect formats and metadata details to be used in setting 
up the pilot of the database prototype.

\section{Database, Interfaces, Metadata}
\label{db_api_meta}

The design of the database for the CAESAR prototype, starting from considerations driven by the 
collected metadata and the requirements to study the chain of phenomena from the Sun to Earth and 
planetary environments, has to face the (common) metadata and data dis-homogeneity. It also has to 
include a way to store and manage the interconnections of phenomena, following Space Weather events 
from the solar starting point through the interplanetary medium, up to Earth, planetary environments 
and satellites.
The pilot database design has already being done and, on top of that, interface solution and user 
interfaces are under development. 
Users will have two ways to access the database: through a graphical web interface to rapidly discover, 
access, visualise and compare data and/or models, or through ASPIS.py, a python module connected to more 
general heliospheric and solar physics python libraries (e.g. SunPy).

\section{Where from here}
\label{closing}

The CAESAR project has about another year of development to provide a final prototype for the ASPIS 
data infrastructure. In this remaining time span data and metadata ingestion on the pilot prototype 
will happen, together with interfaces setup. After that, based on feedback on usage and lessons 
learned, a subsequent refinement of the design will happen, leading to the final data and 
metadata ingestion and full prototype deployment to be handed back to ASI.

\acknowledgements This research has been carried out in the framework of the CAESAR (Comprehensive 
spAce wEather Studies for the ASPIS prototype Realization) project, supported by the Italian Space 
Agency and the National Institute of Astrophysics through the ASI-INAF n.2020-35-HH.0 agreement for
the development of the ASPIS (ASI Space weather InfraStructure) prototype of scientific data centre 
for Space Weather.

\bibliography{P22}  


\end{document}